# RESEARCH
**Open Access**

# Linear normalised hash function for clustering gene sequences and identifying reference sequences from multiple sequence alignments

Manal Helal[1,2], Fanrong Kong[2], Sharon C-A Chen[1,2], Fei Zhou[1,2], Dominic E Dwyer[1,2], John Potter[3] and Vitali Sintchenko[1,2*]

## Abstract

**Background:** Comparative genomics has put additional demands on the assessment of similarity between sequences and their clustering as means for classification. However, defining the optimal number of clusters, cluster density and boundaries for sets of potentially related sequences of genes with variable degrees of polymorphism remains a significant challenge. The aim of this study was to develop a method that would identify the cluster centroids and the optimal number of clusters for a given sensitivity level and could work equally well for the different sequence datasets.

**Results:** A novel method that combines the linear mapping hash function and multiple sequence alignment (MSA) was developed. This method takes advantage of the already sorted by similarity sequences from the MSA output, and identifies the optimal number of clusters, clusters cut-offs, and clusters centroids that can represent reference gene vouchers for the different species. The linear mapping hash function can map an already ordered by similarity distance matrix to indices to reveal gaps in the values around which the optimal cut-offs of the different clusters can be identified. The method was evaluated using sets of closely related (16S rRNA gene sequences of *Nocardia* species) and highly variable (VP1 genomic region of Enterovirus 71) sequences and outperformed existing unsupervised machine learning clustering methods and dimensionality reduction methods. This method does not require prior knowledge of the number of clusters or the distance between clusters, handles clusters of different sizes and shapes, and scales linearly with the dataset.

**Conclusions:** The combination of MSA with the linear mapping hash function is a computationally efficient way of gene sequence clustering and can be a valuable tool for the assessment of similarity, clustering of different microbial genomes, identifying reference sequences, and for the study of evolution of bacteria and viruses.

## Background

The exponential accumulation of DNA and protein sequencing data has demanded efficient tools for the comparison, analysis, clustering, and classification of novel and annotated sequences [1,2]. The identification of the cluster centroid or the most representative [voucher or barcode] sequence has become an important objective in population biology and taxonomy [3-5]. Progressive Multiple Sequence Alignment (MSA) methods perform tree clustering as an initial step before progressively doing pair-wise alignments to build the final MSA output. For example, MUSCLE MSA [6] builds a distance matrix by using the *k-mers* distance measure that does not require a sequence alignment. The distance matrix can then be clustered using the Unweighted Pair Group Method with Arithmetic Mean (UPGMA) [6]. MUSCLE iteratively refines the MSA output over three stages to produce the final output. Evidence suggests that the MUSCLE MSA output outperforms T-COFFEE and ClustalW, and produces the higher Balibase scores [7,8]. Unsupervised machine learning methods such as hierarchical clustering (HC)

* Correspondence: vitali.sintchenko@swahs.health.nsw.gov.au
[1]Sydney Emerging Infections and Biosecurity Institute, Sydney Medical School - Westmead, University of Sydney, Sydney, New South Wales, Australia
Full list of author information is available at the end of the article or partitioning methods have been successfully applied for gene sequence analysis [9,10].

Agglomerative HC often employs linkage matrix building algorithms that usually have a complexity of O($n^2$), for example the Euclidean minimal spanning tree [11]. Some linkage algorithms are geometric-based and aim at one centroid (e.g., AGglomerative NESting or AGNES) [12], while others (e.g., SLINK) rely on connectivity graph methods producing clusters of proper convex shapes [13]. Other agglomerative HC methods produce curved clusters of different sizes by choosing cluster representatives (e.g., the CURE algorithm) [14], but are also insensitive to outliers. More advanced graph-based HC algorithms, such as CHAMELEON, handle irregular cluster shapes by using two stages: a graph-partitioning stage utilising the library HMETIS, and an agglomerative stage based on user-defined thresholds [15]. In contrast, divisive HC algorithms often use Singular Value Decomposition by bisecting data in Euclidean space by a hyper-plane that passes through data centroids orthogonally to an eigenvector with the largest singular value, or the largest $k$-singular values (e.g., Principal Direction Divisive Partitioning) [16]. However, many clustering methods work well only with sequences of high similarity whilst others require training datasets containing known clusters with many members. Nevertheless, defining the optimal number of clusters, cluster density and cluster boundaries for collections of sequences with variable degrees of polymorphism remains a significant challenge [5,17].

Partitioning clustering techniques employ an iterative optimization heuristic greedy technique to gradually improve cluster's coherency by relocating points, which results in high quality clusters. Partitioning clustering techniques can be probabilistic or density-based. The first assume that the data is sampled independently from a mixture model of Bernoulli, Poisson, Gaussian or log-normal distributions. These methods randomly choose the suitable model and estimate the probability of the assignment of the different points to the different clusters. The overall likelihood that the training dataset belongs to the chosen mixture model is calculated by a log-likelihood method, and then Expectation Maximization (EM) converges to the best mixture model. These algorithms include SNOB [18], AUTOCLASS [19], and MCLUST [20]. $K$-means is defined as an algorithm that partitions the dataset into $k$ clusters by reducing the within-group sums of squares using the randomly chosen $k$ centroids representing the weighted average of the points in each cluster [21]. Choosing the optimal $k$ is based on the computationally costly independent running of the algorithm for different $k$ to choose the best $k$. Modifications, such as X-Means, can accelerate this iterative process [22]. Clustering algorithms that utilise density-based partitioning require definitions of density, connectivity and boundary based on a point's nearest neighbours. These methods discover clusters of various shapes and are not sensitive to outliers. Algorithms like DBSCAN, GDBSCAN, OPTICS and DBCLASD relate density to a point in the training data set and its connectivity, while algorithms like DENCLUE relate the density of a point to its attribute space [10]. Other Grid-based methods, such as CLIQUE [23] or MAFIA [24], employ space partitioning rather than data partitioning. However, the dimensionality curse has been a major problem for clustering algorithms as performance degrades significantly when the dimensions (attributes) exceed 20 making the majority of methods described computationally unaffordable for biomedical researchers.

The process of phylogenetic classification of variable-length DNA fragments requires the establishment of "reference sequences" or "DNA barcodes" for species identification and recognition of intra-species sequence polymorphisms or "sequence types" [25,26]. Evidence suggests that such a process of curation, in which the designated most representative sequence of a species (or the "centroid" sequence) is derived from discrete "species groups" of sequences, can be automated (e.g., Integrated Database Network System SmartGene), and improves the species-level identification of clinically relevant pathogens [27,28].

The aim of this study was to develop a clustering method that would work equally well for different sequence datasets as well as identify the cluster centroids and the optimal number of clusters for a given sensitivity level. The method was aimed at optimizing the clustering results both in terms of the mathematically optimal number of clusters and the optimal cluster boundaries.

## Results
### Clustering of gene sequences
Using the Multiple Sequence Alignment (MSA) output in the aligned order (rather than the input order), the sequences are sorted based on the tree building algorithm used, making the closer family of sequences in order before starting another family branch. The MSA is then used to generate a pair-wise distance matrix between the sequences. The produced sorting order made the main diagonal in the distance matrix to be the distance between a sequence and itself (Figure 1). The second diagonal represented the distance between a sequence and its closest other sequence; the third diagonal represented the distance between the sequence and its second closest sequence and so forth. The heat maps in Figure 1 illustrate the rectangular shapes of the darkest blue shades along the diagonal as the boundaries around which the natural selection of a cluster should be identified. The process of identifying these boundaries is based on the linear mapping of the second diagonal values to a normalized index value. The linear mapping to index values employs a deterministic hash function. The hash function used is uniform in the context of the input distance matrix. Thus, a very similar dataset should produce similar or very close hash codes, rather than a highly variable dataset.

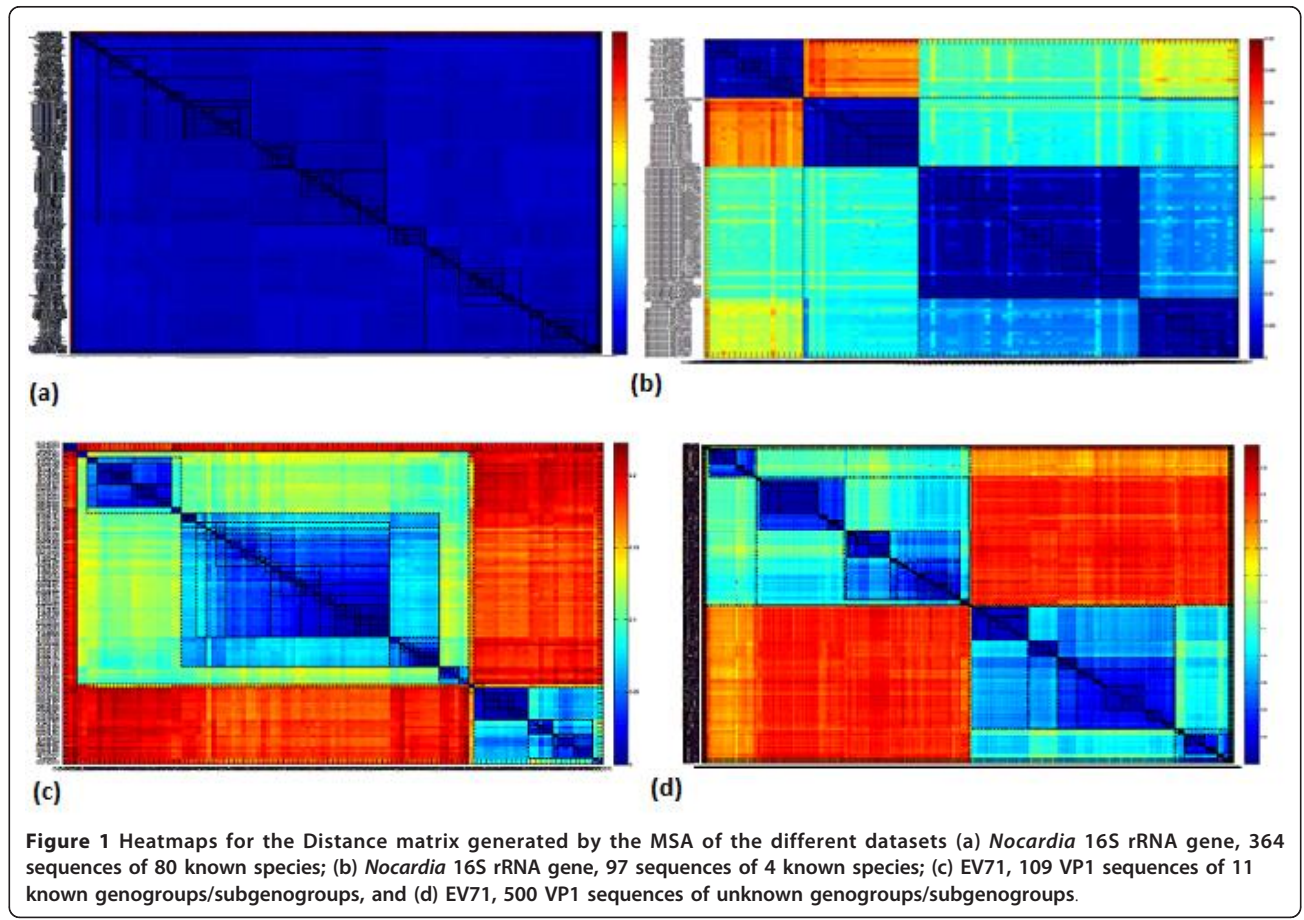

**Figure 1** Heatmaps for the Distance matrix generated by the MSA of the different datasets (a) *Nocardia* 16S rRNA gene, 364 sequences of 80 known species; (b) *Nocardia* 16S rRNA gene, 97 sequences of 4 known species; (c) EV71, 109 VP1 sequences of 11 known genogroups/subgenogroups, and (d) EV71, 500 VP1 sequences of unknown genogroups/subgenogroups.

According to the two sensitivity parameters used (hash range and number of hash codes within one cluster), variable cluster boundaries are shown in Figure 1. These variable boundaries will divide any dataset from a minimum number of clusters up to a maximum number of clusters for each dataset, where neither less nor more divisions can be mathematically feasible. Some highly variable datasets produced a minimum of 16 clusters and a maximum of 29 clusters when a hash range of 4 is used and only one code per cluster, which proves that the distances calculated in the distance matrix control the algorithm output and the minimum and maximum number of clusters will vary from one dataset to another. The number of clusters for both datasets used in this study is plotted in Figure 2 as per the two sensitivity parameters. These variable clustering boundaries can be interpreted equally as abstractness levels (the sensitivity of the clustering in order to decide the optimal number of clusters for the dataset), or as hierarchical levels (encapsulations of sub-clusters within larger clusters).

### Classification of sequences with different degrees of polymorphism

The algorithm was applied to sequences of Enterovirus 71 (EV71) VP1 region (500 sequences), representing polymorphic sequences of highly divergent microorganisms, and 364 sequences of the highly conserved 16S rRNA gene of bacteria from the genus *Nocardia*. The 'heat maps' produced by the Matlab© image function for the distance matrix accurately contrasted the highly recombinant viruses with an over-classified collection of 80 species of *Nocardia* that were less polymorphic and taxonomically closely related. Figure 1 illustrates the combined matrices of similarity measures between sequences of *Nocardia* species (A and B) and EV71 (C and D). The matrices indicate the different borders around the clusters of variable sensitivities around the diagonal (Figure 1).

The identification of the optimal number of clusters was based on the hash range and the number of hash codes (indices) to include in one cluster. Figure 2 shows the number of clusters as per the different Hash ranges on the x-axis and indicates between one and four indices per cluster in the different data series for (a) the 364 sequences of 16S rRNA gene of *Nocardia* of 80 species, (b) the 97 sequences of 16S rRNA gene of *Nocardia* of 4 species, (c) the 109 EV71 sequences with known genogroups/subgenogroups, and (d) the 500 sequences of EV71 VP1 regions. Plots in Figure 2 demonstrate the maximum number of clusters that can be generated for each dataset, after which changing the sensitivity parameters of the hashing function will not divide data into further clusters because of the closeness of the similarity measures.

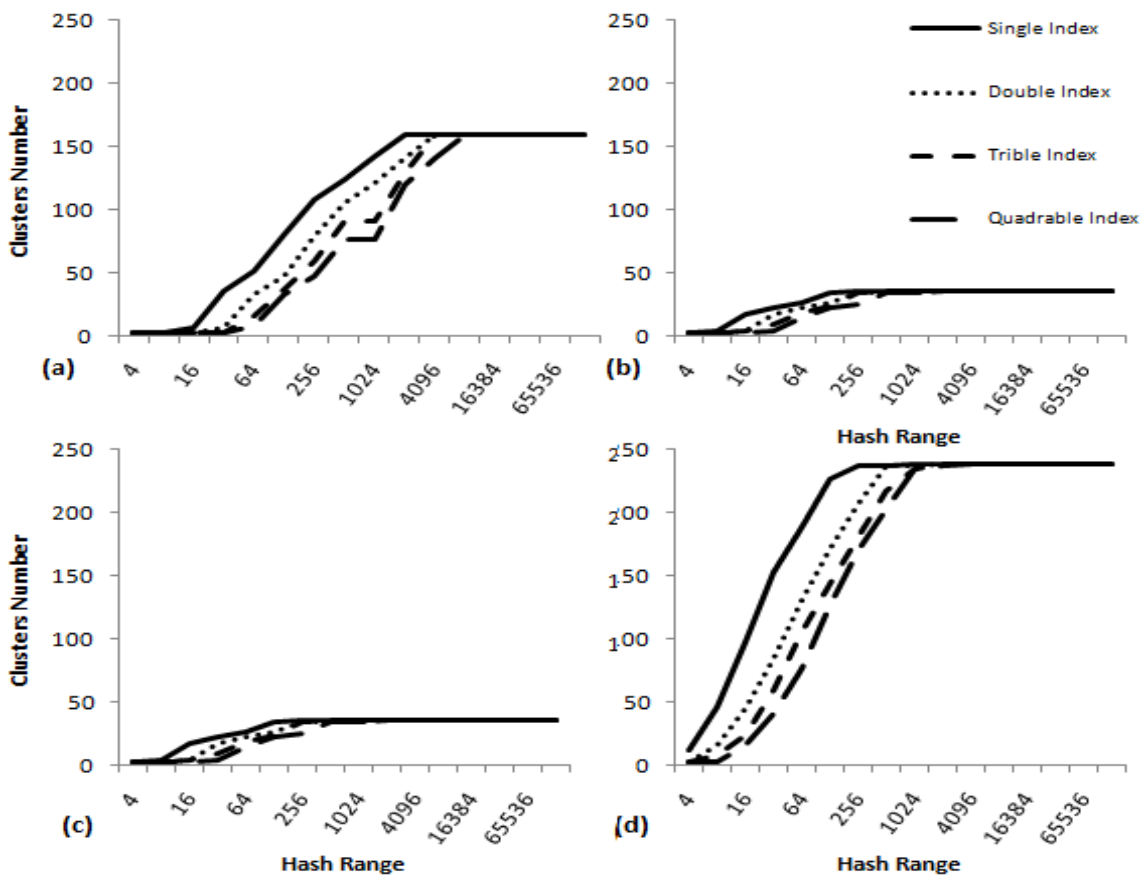

Figure 2 The optimal number of clusters for the different hash ranges and different number of indices per cluster for (a) *Nocardia* 16S rRNA 364 sequences of 80 known species; (b) *Nocardia* 16S rRNA 97 sequences of 4 known species; (c) EV71 109 sequences of 11 known genogroups/subgenogroups; and (d) EV71 500 VP1 sequences of unknown genogroups/subgenogroups.

### Identification of cluster centroids

The combination of MSA with the linear mapping hash function identified the centroid for each cluster, which was then used as a reference. Centroids were the data elements or points positioned in the middle of the cluster cloud, i.e. these were the points with the minimum total distance between them and other points in the same cluster. For a high dimensional dataset, other methods (such as k-means) identify different centroids for each parameter taken as the main parameter, and it is up to the user to decide which parameter should be taken as the basis for the centroid definition. Figure 3 illustrates the PCA plot for the 97 *Nocardia* sequences of four species (*N. cyriacigeorgica*, *N. farcinica*, *N. abscessus*, and *N. nova*). The labels for cluster centroids are shown close to the centroid point highlighted in a square block. Since the clustering and the centroid identification are performed using the linear mapping method, and the PCA plot values are calculated using the PCA scores for the different coordinates (PC1 and PC2 in our case), the centroids do not necessarily fall in a geometric central point in the clusters' two dimensional space.

The implementation of our method produces mean, median, and standard variation distance measures for each cluster. If the training dataset is already composed of a known number of classes (bacterial species or viral genotypes [genogroups/subgenogroups] in our case), the closest number of clusters should represent the optimal one. Otherwise, the optimal sensitivity parameters can be identified by multiple runs and the selection of parameters that minimize the standard variation of intracluster similarity. The PCA plot for the first dataset of 110 EV71 sequences is presented in Figure 4 and illustrates clusters of different sizes and shapes, demonstrating that the method is not likely to be sensitive to outliers. Centroids do not necessarily lie in a geometric middle position in the cluster because their calculation method is based on the distance sub-matrix of the cluster and not on the principal component transformation which is used for the two-dimensional plot.

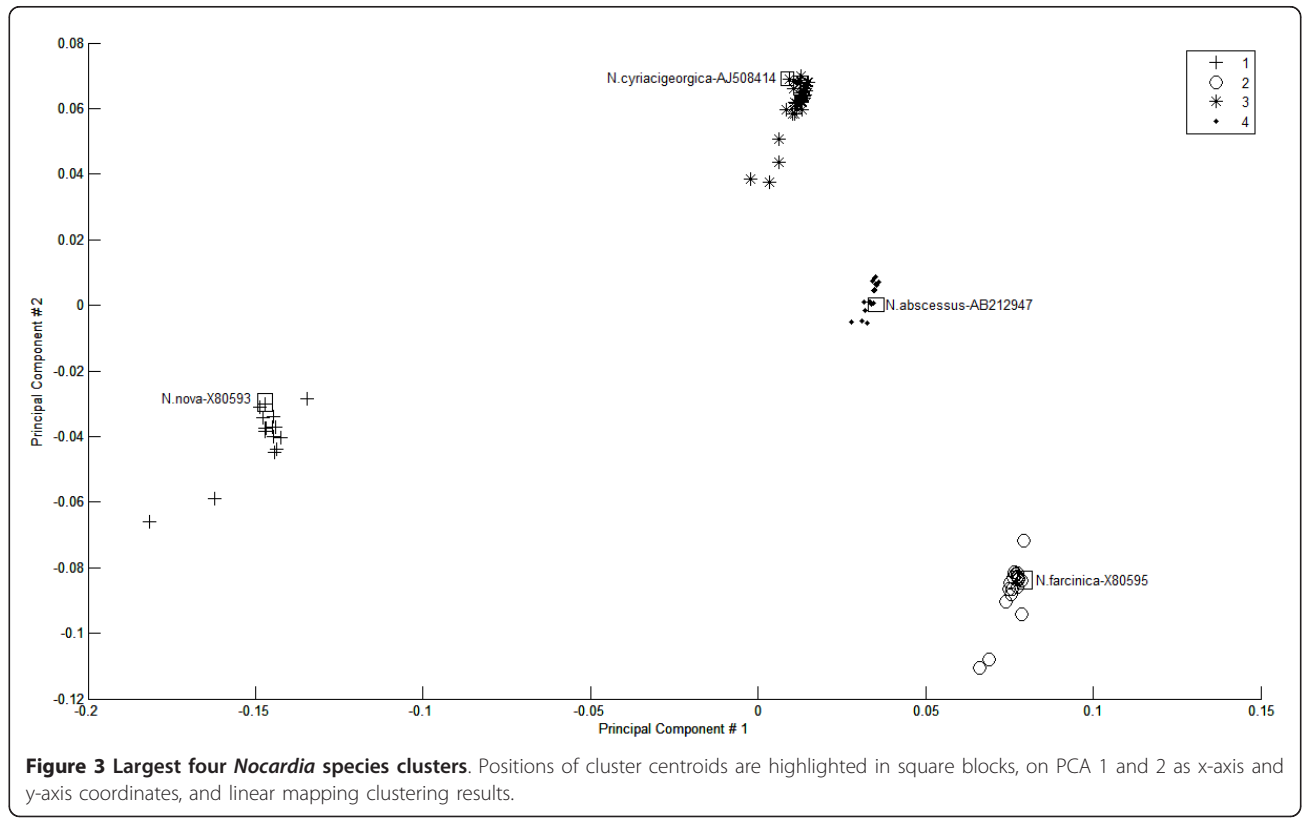

**Figure 3 Largest four *Nocardia* species clusters**. Positions of cluster centroids are highlighted in square blocks, on PCA 1 and 2 as x-axis and y-axis coordinates, and linear mapping clustering results.

## Discussion

This paper described a new method for the efficient optimal clustering of gene sequences and the identification of the most representative members of each cluster from the alignment of multiple sequences. This method identifies discontinuities between each group of pairwise distances to delineate cluster boundaries and to define the most "optimal" clusters. It applies Principal Component Analysis as a dimensionality reduction to represent pair-wise distances between a dataset of aligned sequences [29]. The method sorts the data points (sequences) by shortest distances along the second diagonal of the distance matrix and defines the optimal cluster cut-offs among this diagonal. The method utilises a binary tree-building algorithm as linear sorting algorithms could not produce the required output for two-dimensional data. While other unsupervised clustering methods require either a known number of clusters or a fixed distance between clusters [30], our algorithm defines the optimal number of clusters using a hash function with two sensitivity parameters. The first one is a positive maximum number of indices (hash range) to map the distance measures. The larger the hash range, the bigger the number of clusters that can be generated. The second sensitivity parameter is the number of hash codes to include in one cluster. The more hash codes to encapsulate in one cluster, the fewer the number of clusters produced. The second sensitivity parameter defines an intermediate level of sensitivity before increasing the hash range.

The optimal number of clusters can be defined through data presentation requirements. The analogy of cities, countries and continents can be used, with a city (sequence) described as belonging to a specific country or a continent (parameters). By changing the parameters, the resulting sequence membership to clusters will vary between a larger encapsulating cluster (continent) and a smaller subset of a cluster (country), with bounds on the maximum and minimum of mathematically feasible numbers of clusters. The larger and smaller cluster boundaries maintain the same distance measure information, so that outliers will always appear in their own clusters relative to the distance measure normalizations and the abstractness level required. This avoids the geometric shape approximations found in other methods, and the averaging found in UPGMA. Increasing the sensitivity parameters in the cities, countries and continent analogy, with the city falling on the right edge of a country which in turn falls on the right edge of a continent A that has a neighbouring continent B from the right side, means that the city will continue belonging to the encapsulating continent A only when the calculated distance between this city and the nearest neighbouring city in continent A is shorter than the distance between this city and the nearest neighbouring city in continent B. If the city falls in its own country and continent and the distance between it and its neighbouring city in another continent remains larger than the distances between the second city and other cities in the other continent, then the first city will remain in its own continent in the higher level of abstractness.

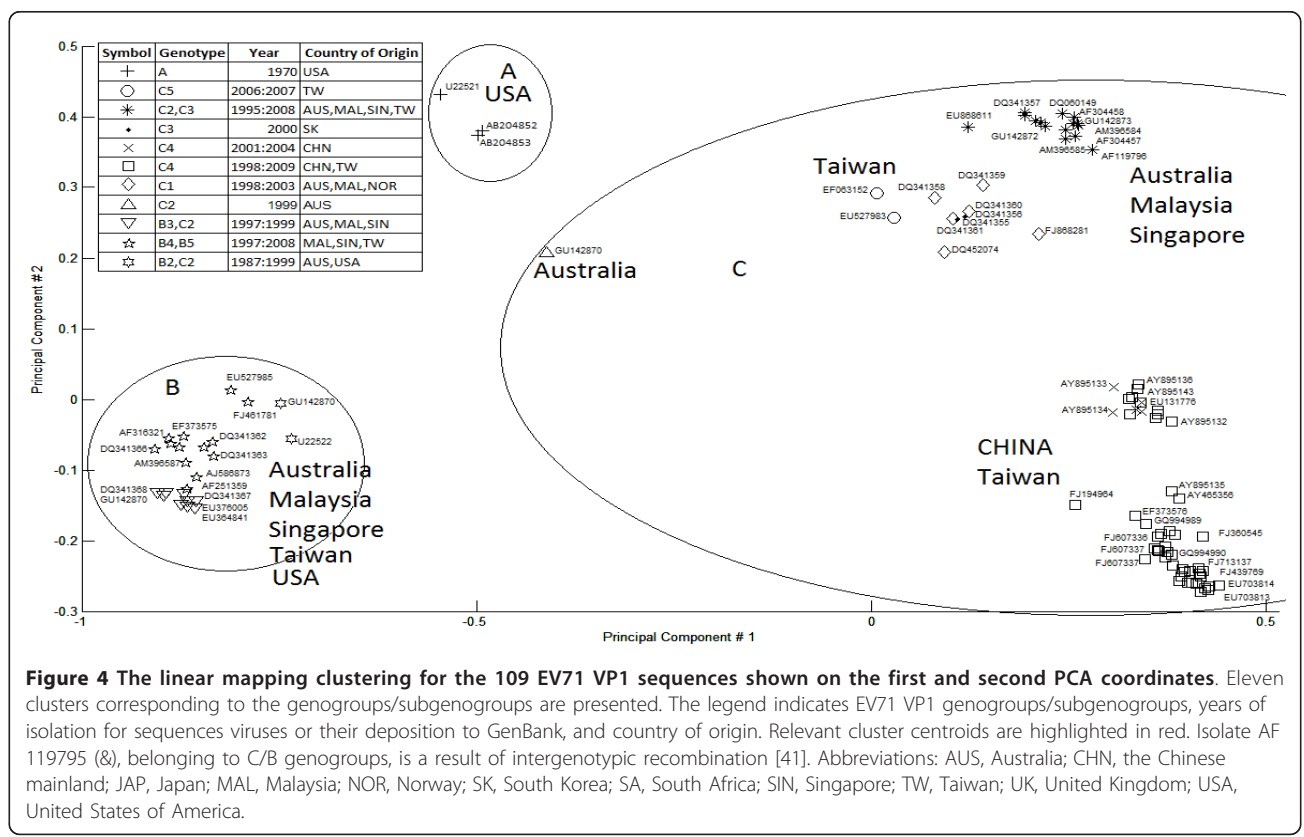

**Figure 4 The linear mapping clustering for the 109 EV71 VP1 sequences shown on the first and second PCA coordinates**. Eleven clusters corresponding to the genogroups/subgenogroups are presented. The legend indicates EV71 VP1 genogroups/subgenogroups, years of isolation for sequences viruses or their deposition to GenBank, and country of origin. Relevant cluster centroids are highlighted in red. Isolate AF 119795 (&), belonging to C/B genogroups, is a result of intergenotypic recombination [41]. Abbreviations: AUS, Australia; CHN, the Chinese mainland; JAP, Japan; MAL, Malaysia; NOR, Norway; SK, South Korea; SA, South Africa; SIN, Singapore; TW, Taiwan; UK, United Kingdom; USA, United States of America.

MUSCLE produces its own distance matrix using the $k$-mers method. The evolutionary distance between each pair of sequences is estimated by computing the fraction of common $k$-mers (substrings of length $k$) in a compressed amino acid alphabet. UPGMA is then used to build the tree topology [6]. However, it does not print it in the output to the user. Therefore, a regeneration of the distance matrix is done from the produced MUSCLE alignment using the distMat application from the EMBOSS suite [31] in some experiments, or the DnaDist application from the Phylib package [32] in other experiments. Euclidean distances using Matlab functions are also evaluated [33]. Importantly, there is a maximum number of clusters that can be generated for every dataset, and once this maximum is reached, increasing the hash range will not increase the maximum possible number of clusters. The hashing function will not be able to merge the dataset into fewer clusters if the distances between the data points provide large gaps that cannot be over-fitted into one cluster. And similarly, it is not possible to divide a cluster into more clusters if all data points included in this cluster are too close to each other, as compared to other data points in other clusters and in the whole dataset. This normalization does not lose the original distance information - it is only translateed into a cluster membership.

The quality of the alignment and the distance matrix calculation method can affect the clustering results

[8,34]. If the alignment order has already been performed in the MUSCLE MSA output, and is preserved in the distance matrix, a classical exact clustering algorithm would sort the data elements, which repeats steps that are already included in the MUSCLE method. The important unanswered questions in this instance revolve around the optimal number of clusters for the dataset, the optimal cut-offs (boundaries) for these clusters, and the choice of reference vouchers for species (cluster representatives from the datasets) [35]. Our approach takes advantage of the information already included in the MSA output, and focuses on optimizing the clustering results. It captures and visualises the complexity of inter- and intra-species relationships and represents gene distances reflecting the discontinuity between species and genogroups (Figures 3 and 4).

The presented method has achieved higher scores than HC, partitioning, and dimensionality reduction clustering methods. In addition, our method does not appear to be sensitive to outliers. Our experience demonstrated that the spatial clustering that relied solely on the PCA plot (like encapsulating the nearby sequences into one cluster) did not produce accurate clustering that matched the existing knowledge about the species [28]. The clustering structure in the dataset might be captured in any of the highest scoring coordinates, not necessarily the first and second, and is often difficult to correlate to other clustering methods. On the other hand, the PCA spatial clustering usually suggests a smaller number of clusters, and is not as sensitive as the presented linear mapping method.

This study adds to the growing range of applications for classification of variable-length DNA fragments. In contrast to many recently proposed tools that target high-volume metagenomics data sets (e.g., Compost-Bin [36] or Phylo-Pythia [37]), our approach utilises relatively small sets of sequences of house-keeping genes in order to identify the most representative sequences. The method described below is deterministic and computationally not expensive. However, it requires the efficient generation of MSA in the alignment order and the distance matrix. The method runs in $O(N)$, where N is the number of sequences, to scan the second diagonal distance matrix values and to identify the cut-offs of the clusters. The dimensionality scalability in this problem's context is still expressed in the number of sequences used, as it clusters the pair-wise distance measures from each sequence to all others in the dataset. The effectiveness of the method can be limited by the performance of the MSA algorithm. For example, the performance of MUSCLE (the MSA method used in this study) for default parameters was defined as $N^4 + NL^2$, where N is the number of sequences, and L is the average sequence length. It was reported that MUSCLE computed an MSA for 5000 sequences with average lengths of 350 bp in 7 minutes [6]. However, the assessment of our method in relation to different distance definitions and comparison with other methods of estimation of the number of clusters, such as UPGMA and the Eigen values, were beyond the scope of this study. These questions warrant further separate investigations. It is noteworthy that UPGMA assumes that a constant rate of evolution is maintained on all sequences from the root. If this assumption fails, the averaging of matrix distances can lead to errors in the topology of results. In such a scenario, the application of Eigen values to estimate the number of clusters based on the UPGMA branch lengths might not be appropriate to define the correct cluster borders (cut-off points). Our approach counterbalances this potential averaging drawback by ignoring the averaged distances (branch lengths) generated by the UPGMA, using the UPGMA ordering of the sequences only (grouped by smallest distance measures), and employing the original distance matrix to define the clusters.

## Conclusions

The combination of MSA with the linear mapping hash function is a computationally efficient way of gene sequence clustering and can be a valuable tool for the assessment of similarity or for the classification of different microbial genomes and for the study of evolution of bacteria and viruses. A linear mapping hash function can map an already ordered by similarity distance matrix to indices to reveal gaps in the values around which the optimal cut-offs of the different clusters can be identified. This method does not require prior knowledge about the number of clusters or the distance between clusters. It is not sensitive to outliers, handles clusters of different sizes and shapes, and scales linearly with dataset of different complexity.

## Methods

### Data sets

Two sets of gene sequences representing distinct taxonomic ranks within viruses and bacteria were used. The first sequence set contained RNA sequences of VP1 genomic regions of enterovirus 71 (EV71). In the first experiment, the similarity of 109 EV71 sequences in VP1 genomic regions classified into three genogroups A, B (including subgenogroups B1, B2, B3, B4, and B5), and C (including subgenogroups C1, C2, C3, C4, and C5) was explored. The alignment was done for 84 complete genomic sequences (average 7415 nucleotides, with min 7312 nucleotides and max 8170 nucleotides), 20 partial genomic sequences containing VP1 regions (average 3234 nucleotides, with min 3202 nucleotides and max 3486 nucleotides), and 5 VP1 regions (all with min 891 nucleotides). In the second experiment, 500 EV71 VP1 sequences with unknown genogroups/subgenogroups (average length 875 nucleotides, with min 410 nucleotides and max 891 nucleotides) were aligned to test the effectiveness of the method on larger datasets of high variability.

The second data set consisted of 364 sequences of the full length 16S rRNA gene of *Nocardia* species downloaded from the GenBank [38], as well as sequences of isolates previously identified in the Clinical Mycology Laboratory, Centre for Infectious Diseases and Microbiology, Westmead Hospital, Sydney [39]. The dataset represented 80 different *Nocardia* species with a relatively low level of polymorphism whose identification was concordant with those archived within the "List of Prokaryotic names with Standing in Nomenclature" [40]. Another experiment was conducted by extracting the sequences of four of the most common *Nocardia* species (*Nocarida cyriacigeorgica*, *Nocardia farcinica*, *Nocardia abscessus*, and *Nocardia nova*) represented by 97 sequences contained in [40].

### Data matrices

Two types of data can be supported: (1) a square data matrix (the number of rows was equal to the number of columns) represented symmetric distance measures of pair-wise distances of data elements, which were listed on the rows and on the columns; and (2) a rectangular matrix containing columns of parametric measures for the data elements as rows. In case of a rectangular data matrix, distances (Euclidian by default) need to be calculated between the parameter variables for every pair of the data elements, and a distance matrix will be generated accordingly.

### Data sorting

Progressive MSA methods rely on tree-building algorithms that cluster the sequences so that the most similar sequences are grouped together. For instance, the MUSCLE method [6] uses the *k-mers* distance measure to build a tree using UPGMA clustering. Consequently, distance matrix generation programs preserved the clustering order on rows and symmetrically on columns. The linear mapping to index values employed a deterministic hash function. The hash function used was uniform in the context of the input distance matrix. Thus, a very similar dataset produced similar or very close hash codes, rather than a highly variable dataset.

The image function implemented in Matlab© [33] mapped the distance values to a colour index, a blue/red colour map of 64 colour shades. The smaller the distance (i.e. highest similarity), the closer the colour shade was to the darkest blue shade. The larger the distance (i.e. highest dissimilarity), the further away the colour index was and the closer it was to the darkest red shade. Accordingly, the clustering process was visualized as the identification of any colour discontinuity on the second diagonal in the distance matrix as a break point to start a new cluster, creating rectangular regions around the diagonal of a single-colour intensity. The linear scaling function in Matlab© was as follows:

```
index = (C-min)/(max-min)*(m-1))+1
```

where C was the distance matrix value, m was the length of the colour map (which was 64 colours by default), and the colour index in the colour map ranged from 1 (darkest blue) to 64 (darkest red). min and max represented the smallest and largest values of the distance matrix.

Using the same mapping function as the clustering hash function, the clustering algorithm generalized to any colour map size (hash range). Every two consecutive index values (as mapped from the diagonal scores in the distance matrix) were compared, and break points that separated the clusters were identified based on the difference. If the difference was greater than the number of indices allowed in one cluster, a new cluster breaking point was selected.

### Multiple sequence alignment

Sequences of microbial genes were sorted according to their unaligned similarity scores by progressive MSA. The MUSCLE MSA algorithm [6] was employed through the EMBL portal, and the final distances were calculated using the *distMat* application from the EMBOSS suite [31] or the DnaDist application from the Phylib package [32]. Both applications take the MSA as input and calculate a distance matrix.

**Hash Function and Clustering**. Using the Matlab programming notation, the following pseudo-code describes the steps of defining the clustering break points, and the starting and ending indices of each cluster.

```
cluster (clusteringData, m, hashrange,
indicesinCluster)
diagscore ← diagonal(clusteringData,
1);
minVal ← min (clusteringData);
maxVal ← max (clusteringData);
cNum ← i ← 1;
sCluster(1) ← 1;
while i < m-1
    index1 ← (diagScore(i)- minVal)/
    (maxVal - minVal)*(hashrange -1))+1;
    index2 ← (diagScore(i+1)- minVal)/
    (maxVal - minVal)*(hashrange-1))+1;
```

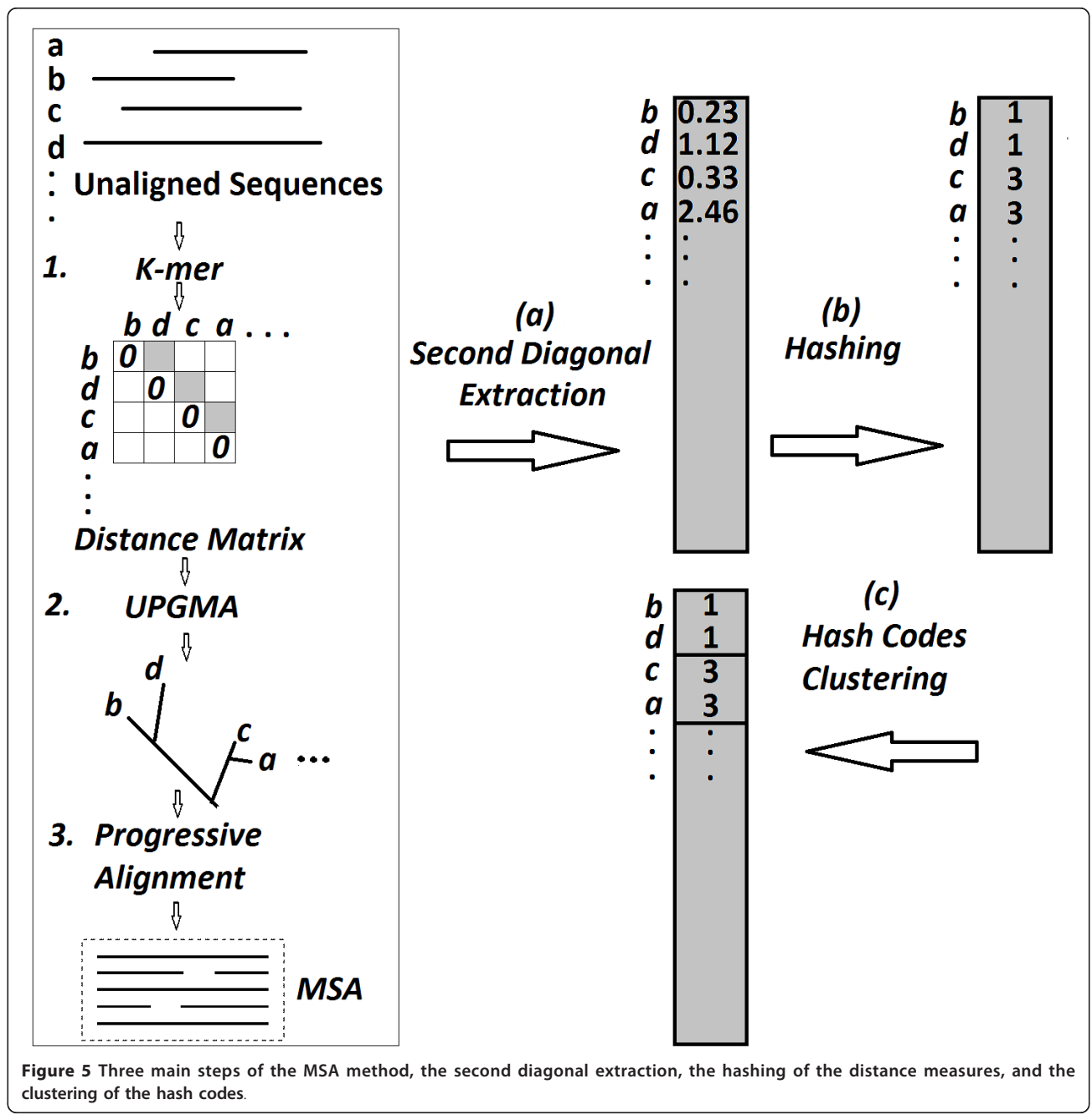

**Figure 5** Three main steps of the MSA method, the second diagonal extraction, the hashing of the distance measures, and the clustering of the hash codes.

```
    if   (abs(index2-index1)   >   =         sCluster(cNum +1) ← eCluster
indicesinCluster)                             (cNum) + 1;
      breakpoint(cNum) ← i + 1;               cNum ← cNum +1;
      if(diagScore(breakpoint(cNum)-1)        i←i+2;
      > diagScore(breakpoint(cNum)))       else
        eCluster(cNum)  ←  breakpoint          i←i+1;
        (cNum) - 1;                          end if
      else                                 end
        eCluster(cNum)  ←  breakpoint      eCluster(cNum) ← m;
        (cNum);                            end cluster
      end
```

The inputs are the "*clusteringData*" (sorted distance matrix), "*m*" (size of the distance matrix, number of sequences), "*hashrange*" (first sensitivity parameter), and "*indicesinCluster*" (second sensitivity parameter). A linear scaling of the "diagScore" uses "hashRange" to determine the scaling and "indicesinCluster" to determine the jump in successive values to determine the edge of a cluster. The algorithm produces the *cNum* (number of clusters) and the *sCluster* and *eCluster* (the starting and ending indices of each cluster). The *diagonal* function returns a vector of the values in the first parameter (*clusteringData*) falling on the diagonal of the order defined in the second parameters. The first diagonal (ranked 0) in the distance matrix should always be equal to zero due to its symmetry. The method retrieves the second diagonal (ranked 1) to cluster based on the distances between a sequence and its most similar following sequence.

**Identification of Centroids**. The sub-matrix for distance measures of a cluster was extracted from the starting and ending indices of this particular cluster (*sCluster* and *eCluster* from the above pseudo-code). If the cluster contains 2 or 1 sequences, the first sequence is considered the centroid. Otherwise, the index of the data element, where the total distance to all other elements in the cluster was the minimum, was then identified. This process is outlined in the following pseudo-code:

```
for i = 1 to cNum
    cSize ← eCluster(i)-sCluster(i)
    if (cSize > 2)
        C_M ← clusteringData(sCluster (i):
        eCluster (i), sCluster (i): eCluster (i))
        for k = 1 to cSize
        distSum(k) ← sum(C_M (k,:))
        end
        [minDist idx] ← min(distSum)
        centroid (i) ← sCluster (i)+idx-1
    else
        centroid(i) ← sCluster (i)
    end
end
```

The main steps of this process are summarised in Figure 5, using an example dataset of four sequences (a, b, c, and d). The MSA method (MUSCLE) builds a distance matrix (*k-mers*) that orders the sequences by similarity as (b, d, c, a), and clusters them by UPGMA as shown in the tree. The remaining steps proposed by this method are the diagonal extraction, the conversion of these scores into hash codes, and the clustering of these codes by identifying the natural gaps between the codes based on the sensitivity parameters chosen.

### Data visualisation

To visualize the spatial distribution of the points on two-dimensional plots, Principal Component Analysis (PCA) was applied on the distance matrix using the Matlab© "princomp" function. The highest scoring first (x-axis) and second (y-axis) coordinates were used for the plotting of the sequences on a two dimensional plane, while the symbols representing the different clusters reflected the clustering results produced by the linear mapping method.

### Acknowledgements
This work was supported by the Australian National Health & Medical Research Council.

### Author details
[1]Sydney Emerging Infections and Biosecurity Institute, Sydney Medical School - Westmead, University of Sydney, Sydney, New South Wales, Australia. [2]Centre for Infectious Diseases and Microbiology, Westmead Hospital, Sydney, New South Wales, Australia. [3]School of Computer Science and Engineering, University of New South Wales, Sydney, New South Wales, Australia.

### Authors' contributions
MH conceived and designed the experiments, carried out experiments and drafted the manuscript. FK participated in the design of the study, in sequence alignment and data analysis. SCAC participated in the design of the study, provided sequencing data and drafted the manuscript. FZ provided sequencing data and participated in the data analysis. DED participated in the data analysis and drafted the manuscript. JP contributed to the development of methods and data analysis. VS conceived and designed the experiments, participated in the data analysis and drafted the manuscript. All authors have read and approved the final manuscript.

### Competing interests
The authors declare that they have no competing interests.